\documentclass[dvips]{article}

\usepackage{icrc2011}
\newcommand{\hess}{HESS~J0632+057}
\newcommand{\lsi}{LS~I~+61~303}

\def\apj{ApJ}%
\def\apjl{ApJL}%
\def\aap{A \& A}%
\def\mnras{MNRAS}%
\title{MAGIC detection of the putative gamma-ray binary HESS~J0632+057}

\newcommand{\etal}{\MakeLowercase{\textit{et al. }}} 
\shorttitle{T. Jogler \etal MAGIC detection of \hess}

\authors{T. Jogler$^{1}$, P. Munar-Adrover$^{2}$, M. Rib\'o$^{2}$, J.~M. 
Paredes$^{2}$ for the MAGIC collaboration}
\afiliations{$^1$Max-Planck-Institut f\"ur Physik, D-80805
M\"unchen, Germany\\ $^2$ Departament d'Astronomia i Meteorologia,
Institut de Ci\`ences del Cosmos (ICC), Universitat de Barcelona
(IEEC-UB), Mart\'{\i} i Franqu\`es 1, E-08028 Barcelona, Spain}
\email{jogler@mppmu.mpg.de}

\abstract{The variable gamma-ray source HESS~J0632+057 is an
excellent candidate for a gamma-ray binary. The putative binary
system was discovered as a point-like VHE gamma-ray source by
HESS. Later measurements by VERITAS yielding no detection,
provided evidence for variable emission in the gamma-ray domain. A
variable X-ray source as well as a Be star (MWC~148) are found at
the location of the gamma-ray source. Recently a periodic X-ray
outburst occurring about every 320 days was reported by
\emph{Swift} (ATel 3152). The putative binary system was observed
by the MAGIC stereo system in 2010 and 2011. Our measurements
demonstrate significant activity in the gamma-ray ($E > 200$ GeV)
band in February 2011. Our detection of the system occurred during
an X-ray outburst reported by \emph{Swift}. Here we present
the obtained light curve and spectrum during this outburst and put
them into context with the X-ray measurements.}
\keywords{gamma-rays, binaries, \hess, MAGIC, VHE, TeV, galactic,
source}

\begin{document}
\maketitle

\section{Introduction}

With the advent of the new generation of Imaging Atmospheric
Cherenkov Telescopes (IACTs) such as MAGIC, HESS and VERITAS, a
new TeV source class, the gamma-ray binaries, could be established.
Only few members of this new class are known. Among these objects
\lsi, LS~5039 and PSR~B1259$-$63 are regularly detected in VHE
gamma rays \cite{MAGIC_lsi_science:2006vk,hess_ls5039,hess_psrb1259}
and all of these three systems show variable or even periodic,
point like VHE (VHE, $E > 100$~GeV) gamma-ray
emission \cite{MAGIC_lsi_periodic:2009ApJ...693..303A,hess_ls5039_period}.
Indeed, all binary systems are expected to be
spatial unresolvable by the current generation of IACTs.\\
A recently discovered unidentified, point-like VHE gamma-ray
source in the constellation of Monoceros by HESS was assumed to be
a gamma-ray binary candidate because of its spatial coincidence
with the Be star MWC~148~\cite{hess_discovery}. The system was
observed by VERITAS in VHE gamma rays from 2006 to 2009 with
sparse sampling and the measurements did not yield any gamma-ray
signal \cite{veritas_ul}. Thus \hess\ is variable in VHE
gamma rays, and as all\footnote{The Crab Nebula is variable in
GeV energies but up to now no TeV variability is measured and thus
it is not counted among the variable VHE gamma-ray sources.} of
the galactic variable VHE gamma-ray sources known today are
associated with binary systems, \hess\ is a very good
binary candidate.\\
Most recently MAGIC and VERITAS detected VHE gamma-ray emission
from the system \cite{magic_atel,veritas_atel}.

Measurements in soft X-rays with \emph{XMM-Newton} indicate an
X-ray source (XMMU J063259.3+054801) at the position of
MWC~148 \cite{hinton_xray}. The source exhibits a power-law
spectrum with spectral index $\Gamma=1.26\pm0.04$ allowing for an
interpretation as synchrotron emission from VHE electrons, although
multi-thermal spectra model yield reasonable fits as well.
Furthermore, the X-ray source showed a variable flux but without
changing the spectral shape. Similar behavior is seen e.g. in the
gamma-ray binary \lsi. Later X-ray measurements conducted with
\emph{Swift} resulted in detecting the same source (identical position) 
but at a different flux level and with a softer
spectral index. The variable nature of the X-ray sources indicate
a binary system. Very recently \emph{Swift} observations from 2009
to 2011 yielded a periodic outburst in the X-ray emission from
\hess\ of $P=321\pm5\textrm{ days}$ \cite{x-ray_outburst}. This is
the best evidence for \hess\ being a gamma-ray binary. \emph{Chandra} 
X-ray measurements during the 2011 Feb X-ray outburst did not find any 
X-ray pulsation in the signal but could demonstrate that the spectrum is 
softer the higher the flux is \cite{X-ray_nanda}.

The region of \hess\ was observed at radio wavelengths, too. The 
measurements conducted in 2008 with the Very Large Array (VLA) and the 
Giant Metrewave Radio Telescope (GMRT) at 5 and 1.28 GHz, respectively, 
detect an unresolved radio source within the position RMS of the VHE 
gamma-ray source and the Be star MWC~148 \cite{radio_counterpart}. The 
source shows variability on the $5\sigma$ level. No extended structures 
were detected at scales of 2 arcseconds, in agreement with the lack of 
such big structures in the other known gamma-ray binaries. During the 2011 
X-ray outburst very high resolution European Very Long Baseline 
Interferometry Network (EVN) observations reveal a point-like source 
coincident with the Be star MWC~148 within 14 milli-arcsec, which evolves 
into an extended source 30 days later \cite{EVN_radio}. The peak of the 
emission is displaced between runs 21 AU, which is bigger than the orbit 
size. The brightness temperature of the source is above $2\times10^6$ K. 
The morphology, size, and displacement on AU scales are similar to those 
found in the other gamma-ray binaries, supporting a similar nature for 
HESS J0632+057 \cite{EVN_radio}.

Optical radial velocity measurements were taken on MW~148, to
verify if it is a member of a binary system~\cite{optical_period}.
The fit to the data yielded a lower limit on the possible period
of the system of $P>100\textrm{ days}$. Very recent optical radial
velocity measurements during the X-ray outburst of 2011 could only
exclude that the X-ray outburst happened during the periastron
passage of the putative compact object
companion~\cite{optical_outburst_atel}. Thus the verification of
the binary nature as well as the determination of the orbital
parameters is still a pending task.

In this proceeding, we present the MAGIC measurements of \hess\
and put them into multiwavelength context.

\section{Observations}

The observations of \hess\ were performed between 2010 Oct and
2011 Mar using the MAGIC telescopes on the Canary island of La
Palma ($28.75^\circ$N, $17.86^\circ$W, 2225~m a.s.l.), from where
\hess\ is observable at zenith distances above 22$^{\circ}$. The
MAGIC stereo system consists of two imaging air Cherenkov
telescopes, each with a 17~m $\O$ mirror. The observations were
carried out in stereo mode, meaning only shower images which
trigger simultaneously both telescopes are recorded. The
stereoscopic observation mode provides a $5\sigma$ signal above
300 GeV from a source which exhibits 0.8\% of the Crab Nebula flux
in 50 hours observation time. Thus the stereo observations are a
factor of two more sensitive than our single telescope
measurements. Further details on the design and performance of the
MAGIC stereo system can be found
in~\cite{MAGIC_stereo_performance}.

\section{Data Analysis}

The data analysis was performed with the standard MAGIC
reconstruction software. Events that trigger both telescopes are
recorded and further processed. The recorded shower images were
calibrated, cleaned and used to calculated image parameters
individually for each telescope. The energy of each event was then
estimated using look up tables generated by Monte Carlo (MC)
simulated $\gamma$-ray events. In another step further parameters,
e.g. the height of the shower maximum and the impact parameter
from each telescope, were calculated. The gamma hadron
classifications and reconstructions of the incoming direction of
the primary shower particles were then performed using the Random
Forest (RF) method~\cite{magic:RF}. Finally, the signal selection
used cuts in the hadronness (calculated by the RF) and the squared
angular distance between the shower pointing direction and the
source position ($\theta^2$). The energy dependent cut values
were determined by optimizing them on a sample of events recorded
from the Crab Nebula under the same zenith angle range and similar
epochs than the \hess\ data. For the energy spectrum and flux, the
effective detector area was estimated by applying the same cuts
used on the data sample to a sample of MC simulated $\gamma$ rays.
Finally, the energy spectrum was unfolded, accounting for the
energy resolution and possible energy reconstruction
bias~\cite{magic:unfolding}.

\section{Results}

We detect VHE gamma-ray emission from the \hess\ data set recorded
in 2011 Feb with a significance of about $6\sigma$ in about 6
hours. A positive detection of the system is only found in the
2011 Feb data which is simultaneously taken to the X-ray outburst
observed by \emph{Swift}. No indication of significant emission is
found in 2010 Dec or 2011 Mar data. Our light curve indicates a
variable VHE gamma-ray source with variability timescales of about
one month. A correlation of the VHE gamma-ray emission with the
periodic X-ray outburst is suggestive but can not be proven since
we only have simultaneous data for the 2011 outburst and the
sampling of the LC is too sparse for individual night correlation
studies. However, it is evident that only in the time of high
X-ray activity the system was detected by MAGIC. In addition, the
LC during the gamma-ray activity shows a constant flux and no
short time (e.g. day timescale) variability.\\
The obtained spectrum is compatible with a simple power law and
the spectral index is compatible with the one previously reported by
HESS. Furthermore the measured flux level is on a similar level as
the previous detections and well above the 2006--2009 VERITAS
upper limits. The same spectral shape and flux level indicate that
the same process might be at work during the HESS detection and
the 2011 MAGIC detection. From our measurements we can infer that
there was an outburst in VHE gamma rays but, due to the sparse
sampling, the duration as well as possible substructures could not
be resolved.

\section{Conclusions}
The presented VHE gamma-ray detection of \hess\ demonstrates that
this source is most likely a new gamma-ray binary. The emission
might be periodic and the gamma-ray activity takes only place
during the 2011 X-ray outburst. The system emits at the same flux
level and with the same energy spectrum whenever detected in VHE
gamma rays, and the detections available up to now are separated four 
years. In case of a periodic modulation with a period of about
320~days, as in X-rays, such
a behavior is well expected.\\
For testing the hypothesis that the VHE gamma-ray and X-ray
emissions are both periodic, future measurements in the two energy
bands are needed. Fortunately, with the X-ray periodicity these
measurements can be planned well in advance.

\section*{Acknowledgments}

We would like to thank the Instituto de Astrof\'{\i}sica de
Canarias for the excellent working conditions at the Observatorio
del Roque de los Muchachos in La Palma. The support of the German
BMBF and MPG, the Italian INFN, the Swiss National Fund SNF, and
the Spanish MICINN is gratefully acknowledged. This work was also
supported by the Marie Curie program, by the CPAN CSD2007-00042
and MultiDark CSD2009-00064 projects of the Spanish
Consolider-Ingenio 2010 programme, by grant DO02-353 of the
Bulgarian NSF, by grant 127740 of the Academy of Finland, by the
YIP of the Helmholtz Gemeinschaft, by the DFG Cluster of
Excellence ``Origin and Structure of the Universe'', and by the
Polish MNiSzW Grant N N203 390834.
M.R. acknowledges financial support from MICINN and European Social Funds 
through a \emph{Ram\'on y Cajal} fellowship.
J.M.P. acknowledges financial support from ICREA Academia.

\clearpage

\end{document}